\documentclass[aps,pra, 10pt,superscriptaddress,twocolumn,floatfix]{revtex4-1}
\usepackage[english]{babel} 
\usepackage{bbm} 
\usepackage{amsmath} 
\usepackage{amssymb} 
\usepackage{physics} 

\usepackage[normalem]{ulem} 
\usepackage{xfrac}
\usepackage{graphicx}
\usepackage[]{xcolor} 

\usepackage{float}

\allowdisplaybreaks

\begin{document}

\title{Controlled-squeeze gate in superconducting quantum circuits}

\author{Nicol\'as F.~Del Grosso} 
\affiliation{ Departamento de F\'\i sica {\it Juan Jos\'e Giambiagi}, FCEyN UBA and IFIBA UBA-CONICET, Facultad de Ciencias Exactas y Naturales, Ciudad Universitaria, Pabell\' on I, 1428 Buenos Aires, Argentina} 
\author{Rodrigo G. Corti\~nas}
\affiliation{Department of Physics and Applied Physics, Yale University, New Haven, CT 06511, USA}
\author{Paula I. Villar}
\affiliation{ Departamento de F\'\i sica {\it Juan Jos\'e Giambiagi}, FCEyN UBA and IFIBA UBA-CONICET, Facultad de Ciencias Exactas y Naturales, Ciudad Universitaria, Pabell\' on I, 1428 Buenos Aires, Argentina} 
\author{Fernando C. Lombardo} 
\affiliation{ Departamento de F\'\i sica {\it Juan Jos\'e Giambiagi}, FCEyN UBA and IFIBA UBA-CONICET, Facultad de Ciencias Exactas y Naturales, Ciudad Universitaria, Pabell\' on I, 1428 Buenos Aires, Argentina} 
\author{Juan Pablo Paz} 
\affiliation{ Departamento de F\'\i sica {\it Juan Jos\'e Giambiagi}, FCEyN UBA and IFIBA UBA-CONICET, Facultad de Ciencias Exactas y Naturales, Ciudad Universitaria, Pabell\' on I, 1428 Buenos Aires, Argentina}

\begin{abstract}
\noindent 
We present a method to prepare non-classical states of the electromagnetic field in a microwave resonator. It is based on a controlled gate that applies a squeezing operation on a SQUID-terminated resonator conditioned on the state of a dispersively coupled qubit. This controlled-squeeze gate, when combined with Gaussian operations on the resonator, is universal. We explore the use of this tool to map an arbitrary qubit state into a superposition of squeezed states. In particular, we target a bosonic code with well-defined superparity which makes photon losses detectable by non-demolition parity measurements. We analyze the possibility of implementing this using state-of-the-art circuit QED tools and conclude that it is within reach of current technologies.

\end{abstract}
\maketitle
\section{Introduction} \label{sec:intro}

Circuit quantum electrodynamics (cQED) has become the leading architecture for quantum computation. With this kind of setup one can design different types of qubits using appropriately chosen combinations of Josephson junctions, capacitors, and inductors controlling them with microwave fields \cite{Blais2021,Google_2022,Krinner_2022}. 
Circuit QED has already been used to manipulate tens of qubits for quantum simulation \cite{houck2012chip,andersen2024thermalization} and quantum error correction \cite{Google_2022, reed2012realization}.
Even at the small scale of a few resonators, cQED provides an alternative to cavity QED \cite{haroche2020cavity} with practical advantages. It has been used to study the dynamical Casimir effect (DCE) \cite{wilson,paraoanu}, foundational aspects of quantum mechanics \cite{Devoret1985,minev2019},  and also enabled practical applications like error correction of non-classical states in resonators \cite{ofek2016,sivak2022} and quantum communication \cite{storz2023loophole}.

The preparation and control of non-classical quantum states of the electromagnetic field in the resonator is crucial to perform universal simulations which in principle involve the manipulation of arbitrary quantum states. This has a long and interesting history which includes the preparation of quantum superpositions of coherent states (Schr\"odinger cat states) using field displacement or rotations controlled by the state of the qubit \cite{monroe1996, brune1996, haroche2006}. These studies are not only interesting from the technological point of view but mostly from their fundamental implications and enabled, for example, the first real-time analysis of the process of decoherence in a cavity QED setup \cite{haroche2006, deleglise2008}.

To generate arbitrary states of the cavity field with limited resources it is crucial to identify a universal set of gates \cite{khaneja2005optimal,leghtas2013deterministic,heeres2017implementing,Krastanov2015, Kundra2022}. There are several techniques for universal control in the dispersive regime, these methods include the qubit cavity mapping protocol (qcMAP) \cite{leghtas2013deterministic}, the Selective Number-dependent Arbitrary Phase (SNAP) and displacement gate set \cite{Krastanov2015,heeres2017implementing,fasel,Kundra2022}, measurement-based methods for oscillator state preparation \cite{wang}, or model-based pulse shaping such as GRadient Ascent Pulse Engineering (GRAPE) \cite{khaneja2005optimal,heeres2017implementing,reinhold}. 
 In this context, it has been shown that arbitrary unitary operators acting on the resonator field states can be generated using Gaussian operations together with controlled displacement operators \cite{eickbusch2022, eickbush-thesis} (which are denoted as $\textbf{C-Dsp}(\gamma)$ and apply a displacement $\hat{D}(\gamma)$ depending on the state of a control qubit). Complemented with Gaussian operations on the field, and single qubit operations, this controlled-gate provides a universal resource to create arbitrary states.

In this paper we present another gate, based on controlled squeezing, that can also be used as part of a universal set. We also show that it can be implemented using reasonable experimental resources. We will denote it by $\textbf{C-Sqz}(r,\theta)$ as it applies the squeezing operation $\hat{S}(r,\theta)$, where $\hat{S}(r,\theta) = \exp(r/2 (e^{-i \theta} \hat{a}^2 -  e^{i \theta} \hat{a}^{\dagger 2}))$, conditioned on the state of a control qubit. We also present an example protocol to make use of this gate in the cQED setup to encode quantum states in the resonator using an encoding that makes the errors induced by the loss of a photon detectable through parity measurement. 

Before describing the way to implement the $\textbf{C-Sqz}(r,\theta)$ gate, it is worth stressing the simplicity of the proof of its universal nature. We will briefly sketch the demonstration here and present a more detailed one in Appendix \ref{universalidad}. In short,  the universality of controlled-squeezing can be shown to be equivalent to the universality of $\textbf{C-Dsp}(\gamma)$. Thus, as shown in  Appendix \ref{universalidad}, when applying squeezing $\hat{S}(r,\theta)$ and  anti-squeezing  operators $\hat{S}^\dagger(r,\theta)$ before and after a displacement $\hat{D}(\gamma)$, we obtain a displacement operator $\hat{D}(\gamma\prime)$, where $\gamma\prime$ depends both on $\gamma$ and the squeezing parameters (this idea was employed by Wineland and co-workers to develop a motional amplifier, using an amplitude-modulated ion trap to generate squeezing \cite{wineland-science,wineland2013}). 

This relation between squeezing and displacement operators (as shown in the appendix) can be generalized to the case of controlled displacement and controlled squeezing gates. In this way, we simply show that the use of an arbitrary $\textbf{C-Sqz}(r,\theta)$ gate, together with a fixed displacement operator, allows the implementation of arbitrary controlled displacement, which is universal. Therefore, the universality of $\textbf{C-Sqz}(r,\theta)$ arises from that of $\textbf{C-Dsp}(\gamma)$.

Choosing a universal set of gates over others is a decision that should be made considering the hardware at hand and the target states or operators desired. 
We simply demonstrate that both universal gates can be implemented within the same setup, adding flexibility to the cQED architecture and simplifying certain tasks.

This article is organized as follows. In the next section (Sec.\ref{sec:setup}), we will present the setup for implementing $\textbf{C-Sqz}(r,\theta)$ in an architecture of cQED. In Sec.\ref{sec:encoding} we will show how to encode an arbitrary qubit state in the resonator in an error detectable way. Sec. \ref{sec:implemantation} will be dedicated to analyse the implementation of the above encoding protocol. Finally, we will summarize our proposal in the Conclusions section \ref{sec: conclusiones}.

\begin{figure}
	\centering	\includegraphics[width=0.5\textwidth]{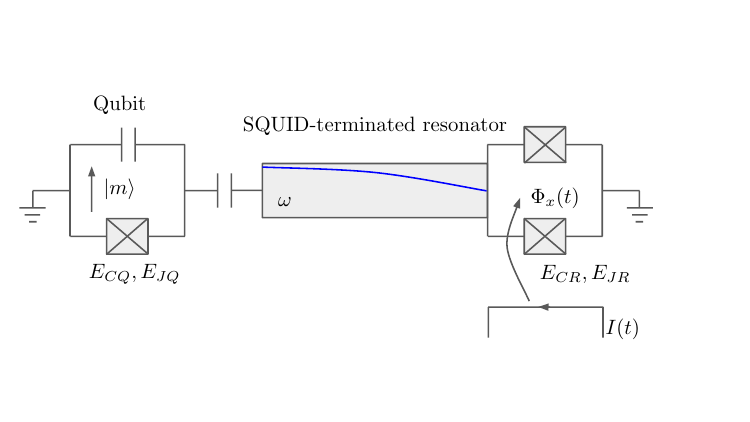}
    		\caption{The setup proposed consists of a coplanar wave guide acting as a resonator and terminated in a SQUID at the right end and capacitively coupled to a transmon at the left. The SQUID is used to parametrically drive the resonator by applying a time dependent magnetic flux to it. The transmon qubit is initially prepared in a quantum state $|m\rangle$ (where $m=0,1$) and acts as a switch that either tunes or detunes the resonator to the driving frequency. Blue solid-line represents the mode within the resonator}
	\label{fig:stup}
\end{figure}

\section{Setup for the implementation of A controlled-squeeze gate}  \label{sec:setup}

To implement the proposed controlled-squeeze gate we use two basic elements: a frequency-tunable resonator and a qubit \cite{delsing-snail,wang2022towards} with quantum states $\vert 0\rangle$ and $\vert 1\rangle$. This qubit is dispersively coupled to the resonator in the number splitting regime \cite{splitting}, so that its frequency depends of the state of the qubit (i.e. it is $\bar\omega_0$ if the state is $\vert 0\rangle$ and $\bar\omega_1$ if the state is $\vert 1\rangle$).
The resonator is terminated by a SQUID, where we apply a time-dependent flux $\Phi(t) = \epsilon\sin(\omega_d t)$ where $\epsilon$ is the amplitude of the flux drive. In this way the resonator's natural frequency becomes a time-dependent parameter $ \omega_{0,1}(t)=\bar\omega_{0,1}+ g_d \sin(\omega_d t-\theta)$,  where $g_d = \bar g_d \epsilon$, with $\bar g_d$ the driving coupling constant between the SQUID and the resonator modes (see below). When the driving frequency is $\omega_d = 2 \bar\omega_1$, then parametric resonance takes place and the cavity's field is squeezed as a result \cite{Squeezing-njp,squeeze_states_param}. If the detuning $\Delta = \bar\omega_1- \bar\omega_0$ is $\Delta\gg g_d$ then the state of the cavity field is effectively squeezed only if the control qubit is in the state $\vert 1\rangle$. On the other hand, if the state is $\vert 0\rangle$, ordinary harmonic evolution with frequency $\bar\omega_0$ takes place and when compensated as described below, turns this into a controlled-squeeze gate.

The setup we envision is described in Fig. 1: the transmon qubit \cite{transmon-2007} on the left is capacitively coupled with a $\lambda/4$ resonator terminated by a flux-tunable SQUID \cite{wilson,lambda4}. This last component has been used to demonstrate the DCE \cite{wilson,paraoanu,svensson_period_2018,wilson_photon_2010}, and has been the subject of thorough investigation \cite{lambda4,Wustmann2013}.
The setup in Fig. 1 can be modeled with the Hamiltonian (we use $\hbar = 1$ thorough the paper)

\begin{equation}
\label{eq:1}
    \hat H (t) =  \frac{\omega_q}{2}\hat{\sigma}_z + \omega \hat a^{\dagger} \hat a  + \chi \hat a^{\dagger}\hat a \hat \sigma_{z} + g_d \sin(\omega_d t-\theta) (\hat a^{\dagger} +\hat a)^2  ,    
\end{equation}
where $\hat a$ is the bosonic annihilation operator of the resonator mode and $\hat\sigma_z=|0\rangle\langle0|-|1\rangle\langle1|$ is the Pauli operator associated with the qubit. Here, $\omega_q$ is the frequency of the qubit, and $\chi$ is the dispersive coupling constant between the resonator and the transmon (that is considered in the two-level approximation in Eq.(\ref{eq:1})). The interaction term between the qubit and the resonator is the result of the one-mode approximation for the resonator and the two-level system for the transmon. Then, within the rotating
wave approximation (RWA) the Jaynes-Cumming type of coupling turns out to be directly $\chi \hat a^{\dagger}\hat a \hat \sigma_{z}$ in the non-resonant regime. The dependence of these constants on physically relevant parameters such as the Josephson energy, the capacitance of the transmon-qubit, its charge energy, the coupling capacitance, the inductance and capacitance per unit length of the resonator and the parameters characterizing the external pumping are discussed in detail in the Appendix \ref{ap:Hamiltonian}.

The Hamiltonian in Eq.(\ref{eq:1}) does not include 
non-linearities, which as discussed in the Appendix \ref{ap:Hamiltonian}, can be neglected when the ratio between the 
inductive energy of the resonator and the Josephson energy of the SQUID,
 ($E_{LR}/E_{JR}$) is small (see labels in Fig. \ref{fig:stup}).
In what follows we will work under this assumption and explain the effect of the Hamiltonian (\ref{eq:1}), presenting later some results with experimentally realizable parameters discussing also the effect of losses, decoherence, and non-linearities. A similar Hamiltonian appears in the context of trapped ions where a gate that squeezes the motional degree of freedom of an ion depending on its internal state was proposed by modulating the amplitude of an optical lattice \cite{paz2020}.

Looking at the Hamiltonian in Eq.(\ref{eq:1}) one can see that it describes a harmonic oscillator with a resonance frequency $ \omega_{0,1}(t)$
that both varies in time and is conditioned on the qubit state ($\bar\omega_0=\omega+\chi$ when the qubit is in the state $|0\rangle$ or $\bar\omega_1=\omega-\chi$ when the qubit is in the state $|1\rangle$). 
If the system is driven with $\omega_d = 2 \bar\omega_1$, and the qubit state is $|1\rangle$, then the parametric resonance is exited and the state of the resonator is squeezed. On the other hand, if the state of the qubit is $|0\rangle$, the resonator simple acquires a renormalized frequency due to the effect of the AC-Stark shift. Thus, in the frame rotating with frequency $\bar\omega_1$, the Hamiltonian (\ref{eq:1}), within the RWA,  reads as    

\begin{equation}
\hat H_{I} = \frac{1}{2}ig_d (e^{-i\theta} \hat a^{2}-e^{i\theta}\hat a^{\dagger2})\otimes |1\rangle\langle1| + \tilde \Delta \hat a^\dagger \hat a \otimes |0\rangle\langle 0|,\label{HI}\end{equation}
where $\tilde \Delta \approx \Delta (1 - 1/2(g_d/\Delta)^{2}) + {\cal O}((g_d/\Delta)^{4})$.

The temporal evolution operator associated with the above Hamiltonian is 

\begin{eqnarray}\exp(-i\hat H_{I} t) &=& \hat S(r,\theta)\otimes |1\rangle\langle1|+ \hat U_0(\varphi) \otimes |0\rangle\langle0|\nonumber \\ &=& \hat U(r, \theta,\varphi),\label{evol}\end{eqnarray}
where $\hat{S}(r,\theta)$ is the squeezing operator defined above, with the squeezing parameter $r=-g_d \, t$ and the squeezing angle $\theta$ set by the phase of the driving. In (\ref{evol}), $\hat U_0(\varphi)$ is the evolution operator of an oscillator with frequency $\tilde \Delta$, which during a time $t$, induces a rotation in phase space in an angle $\varphi = \tilde\Delta t$. For the above evolution operator $\hat U(r,\theta,\varphi)$ in Eq. (\ref{evol}) to be a true controlled-squeeze gate, it is necessary to compensate the free evolution $\hat U_0(\varphi)$. This can be done, at least in two different ways. First, one can  turn off the magnetic driving and then wait a time $\tau$ chosen in such a way that $\Delta \tau + \varphi = 2 k \pi$, for some integer $k$. After this, as $\hat U_0(\Delta \tau) \hat U_0(\varphi) = 1$, the evolution operator is the desired controlled-squeeze gate: $\textbf{C-Sqz}(r,\theta) = \hat U_0(\Delta \tau) \otimes |0\rangle \langle 0| \otimes \hat U(r,\theta,\varphi)$. A different alternative, that's not require  turning off the magnetic driving, is to use the non-compensated controlled-squeeze gate $\hat U(r,\theta,\varphi)$ and to choose the subsequent operations to depend on the rotation angle $\varphi$. We will follow this second strategy in the next section below and we will leave the former case for Appendix \ref{ap: ep}, where we fully detail both encoding protocols.

\section{Encoding protocol}  \label{sec:encoding}

Now we show how to use the above result in order to encode an arbitrary qubit state in a resonator making the errors induced by photons losses detectable \cite{ofek2016,ni2023beating,Grimsmo2020,teoh2023dual}. We choose the encoding in such a way that the logical states $|0\rangle$ and $|1\rangle$  are represented by the states  $|\chi_+\rangle$ and $|\chi_-\rangle$ of the resonator built as even and odd superpositions of states squeezed along two orthogonal directions in quadrature space, i.e.,

\begin{align}
    |\chi_{\pm}\rangle=\frac{1}{\sqrt{2}c_\pm}(|r,\tilde{\theta}\rangle \pm |r,\tilde\theta+\pi\rangle)\label{chipm},
\end{align}
where $|r,\tilde\theta\rangle$ is a one-mode squeezed state and the constant $c_\pm=\sqrt{1\pm1/\sqrt{\cosh 2r}}$.
From this it is simple to see that the states $|\chi_+\rangle$ and $|\chi_-\rangle$ have similar properties to the four-legged cat \cite{ofek2016} states as they are respectively superposition of $4n$ and $4n+2$ photon states, which implies that when loosing a photon the encoded state still stores a coherent superposition and the error can be detected by a subsequent parity measurement of the photon number inside the resonator. This measurement of the cavity, both for parity and for Wigner tomography (which is just the measurement of parity after applying a displacement) can be done using  usual techniques in the field. A sufficient requirement is a dispersively coupled two-level system like the transmon in our proposal.

To prepare a general encoded state, we should start with an arbitrary qubit state $|\Psi_Q\rangle=\alpha|0\rangle+\beta|1\rangle$ and the resonator in the vacuum. Then we apply the following sequence: i) Apply a Hadamard gate to the qubit (transforming $|0\rangle\to(|0\rangle+|1\rangle)/\sqrt{2}$ and $|1\rangle\to(|0\rangle-|1\rangle)/\sqrt{2}$); ii) Apply the non-compensated controlled-squeeze gate $\hat U(r,\theta,\varphi)$ defined in Eq.(\ref{evol}); iii) Apply a $\pi$-rotation to the qubit;  iv) Apply the operator $\hat U(r,\theta + 2 \varphi + \pi, \varphi)$; v) Apply a $\pi$-rotation to the qubit; and vi) Apply another Hadamard gate to the qubit. After this sequence the combined qubit-resonator state will be 
\begin{eqnarray}
    |\Psi_{QR}\rangle &=&\frac{1}{\sqrt{2}}|0\rangle\left(\alpha\ {c}_{+}|\chi_{+}\rangle+\beta\ {c}_{-}|\chi_{-}\rangle\right) \nonumber \\ 
    &+&\frac{1}{\sqrt{2}}|1\rangle\left(\alpha\ {c}_{-}|\chi_{-}\rangle+\beta\ {c}_{+}|\chi_{+}\rangle\right)\label{psiQR}
\end{eqnarray} where $|\chi_{+}\rangle$ and $|\chi_{-}\rangle$ are the above defined ones with $\tilde \theta = \theta + 2 \varphi$. 
If we measure $\hat{\sigma}_z$ for the qubit, we obtain the results $\pm1$, which respectively identify the states $|0\rangle$ or $|1\rangle$, with probability $P_\pm= 1/2 (1\pm P_z/\sqrt{\cosh(2r)})$, where $P_z= \alpha^2-\beta^2$ is the z-component of the polarization vector of the qubit which identifies its state in the Bloch sphere. For each result, the state of the resonator turns out to be
$|\Psi_{R}^\pm\rangle=(\alpha{c}_{\pm}|\chi_{\pm}\rangle+\beta{c}_{\mp}|\chi_{\mp}\rangle)/\sqrt{|\alpha|^2 c_\pm^2  +  |\beta|^2 c_\mp^2}$. 
For each result $\sigma_z = \pm 1$ the resonator stores the encoded states $|\Psi_R^\pm\rangle$ whose fidelity with respect to ideal state $|\tilde{\Psi}_Q^\pm\rangle = \alpha| \chi_\pm\rangle + \beta|\chi_\mp\rangle$ is $F_\pm = |\langle\tilde{\Psi}_Q^\pm|\Psi_R^\pm\rangle|^2$. The average fidelity for the complete encoding protocol is $\bar F = P_+ F_+ + P_- F_-$ which can be expressed as 

\begin{equation} \bar F = \frac{1}{2}(1 + P_z^2) + \frac{1}{2}(1 - P_z^2) \sqrt{1 - \frac{1}{\cosh(2r)}}. \label{averageF}\end{equation} 
The lowest fidelity states are those in the equator, i.e. when $P_z= 0$, and, in the limit of high squeezing, we find that $\bar F\sim 1 - e^{-2r}(1-P_z^2)/4$. Clearly to enforce a high fidelity for every state we need the squeezing factor to be large enough. In fact $\bar F\geq 0.995$ requires $r\geq2$.

\section{Implementation}  \label{sec:implemantation}

We analyzed the implementation of the above encoding protocol considering material properties that have been achieved in systems similar to the one proposed \cite{delsing_new,ganjam2024surpassing}. For this we choose the resonator frequency $\omega/(2\pi)=6 \ \text{GHz}$, the qubit frequency $\omega_q/(2\pi)=4\ \text{GHz}$, the driving coupling $\bar g_d=50 \ \text{MHz}$, the driving amplitude  $\epsilon=0.15$ (which gives $g_d = 7.5 \ \text{MHz}$) and the qubit coupling strength $\chi/(2\pi)=8 \ \text{MHz}$. The controlled-squeeze gate is applied during $200\rm \ n\text{s}$ resulting in a squeezing $r\sim 1.5$. We summarize in Appendix \ref{parametros} all the physical parameters used. 

We included the effect of losses and decoherence modeled through a master equation describing thermal contact between the qubit-resonator system and a bath at $60 \ \text{mK}$. For the relaxation time-scales we choose  values which lying in between the ones reported in Ref. \cite{delsing_new} and the most recent one \cite{ganjam2024surpassing}:  a qubit relaxation time-scale $\tau_q=200 \ \mu\text{s}$, a resonator damping time $\tau_r=200 \ \mu\text{s}$ and a qubit dephasing time-scale of $\tau_\phi=10 \ \mu\text{s}$. 

On chip resonator have been fabricated surpasing the millisecond lifetimes at $5 \ \text{GHz}$. Including the junction for tunability in the resonator will certainly degrade its properties but it is reasonable to expect that the lifetime will remain well above $500 \ \mu\text{s}$ in state-of-the-art Tantalum devices \cite{ganjam2024surpassing}.

The state of the joint system, resonator plus qubit, is then described by a density matrix. Taking a harmonic drive $\delta\phi(t)=\epsilon \Phi_0 \sin(\Omega t-\theta)$ for the external magnetic flux and performing the rotating wave approximation, the density matrix in the interaction picture evolves according to the equation
\begin{align}
\dot{\rho}_{I}(t)&=-\frac{i}{\hbar}[H_{I}(t),\rho_{I}(t)]\nonumber\\
&+\frac{1}{2}\kappa_{1}{\cal D}[a]\rho_{I} + \frac{1}{2}\kappa_{2} {\cal D}[a^\dagger]\rho_{I} + \frac{1}{2}\kappa_{1}^{\prime} {\cal D}[\sigma_-]\rho_{I}\nonumber\\
&+\frac{1}{2}\kappa_{2}^{\prime} {\cal D}[\sigma_+]\rho_{I}
+\frac{1}{2}\kappa_{\phi} {\cal D}[\sigma_z]\rho_{I}\nonumber,
\end{align}
where we have introduced the dissipator ${\cal D}[{\cal O}] \rho = {\cal O} \rho \, {\cal O}^\dagger - 1/2 \, {\cal O} {\cal O}^\dagger \, \rho \, - 1/2 \, \rho \, {\cal O} {\cal O}^\dagger$. We have set coefficients in the mster equation as:
\begin{align}
\kappa_{1}&=(n_{\rm th}+1)\kappa \,\, ; \,\, \,\,
\kappa_{1}^{\prime}=(n_{\rm th,q}+1)\kappa^{\prime} \nonumber \\
\kappa_{2}&=n_{\rm th} \kappa \,\, \,\, \,\,\,\,\,\,\,\,\,\,\,\,\, ; \,\, \
\kappa_{2}^{\prime}=n_{\rm th,q}\kappa^{\prime}\nonumber
\end{align}
where $n_{\rm th}=1/(\exp{\omega/(K_B T)}-1)$, $n_{\rm th,q}=1/(\exp{\omega_q/(K_B T)}-1)$, $\kappa=1/\tau_r$ is the decay rate of the resonator, $\kappa^\prime=1/\tau_q$ being the decay rate of the qubit, and $\kappa_\phi=1/\tau_{\phi}$ the dephasing rate of the qubit. The Hamiltonian in the commutator of the first line is given by Eq.(\ref{HI}). 

 The evolution was performed using QuTiP and the Fock space for the resonator was truncated to a basis of dimension $N=90$. Using this parameters and considering that the qubit has a capacitive energy $E_{CQ} =150 \ \text{MHz}$ the induced Kerr non-linearity is in fact 3 orders of magnitude less than the squeezing term making it negligible.

\begin{figure}
\centering	\includegraphics[width=0.49\textwidth]{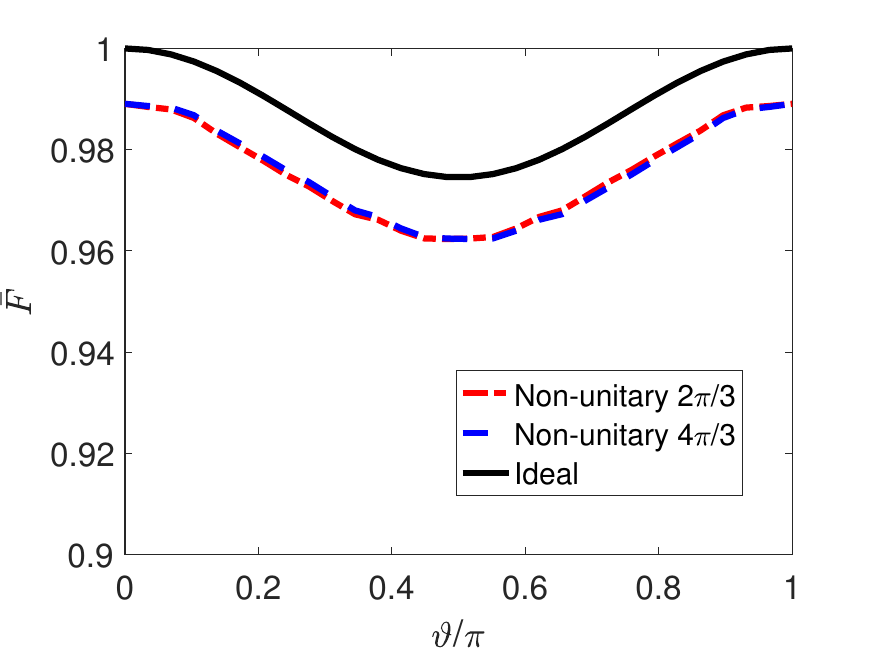}
	\caption{ Average fidelity for encoding a qubit state into the resonator as a function of azimutal angle $\vartheta$ of the Bloch sphere for two different values of polar angle $\phi = 2\pi/3$ and $4\pi/3$. Physical parameters are: $\omega/(2\pi)=6 \ \text{GHz}$; $\omega_q/(2\pi)=4 \ \text{GHz}$; $g_d=7.5  \text{MHz}$; and $\chi/(2\pi)=8 \ \text{MHz}$. The qubit and resonator relaxation time-scales are $\tau_q= \tau_r=200 \ \mu\text{s}$, while the qubit dephasing time is $\tau_\phi=10 \ \mu\text{s}$. Coupling with the thermal bath at $60 \ \text{mK}$ was assumed. Upper solid (black) line corresponds to the ideal case of Eq.(\ref{averageF}) where coupling to the environment is ignored and the optimization process is not taken into account. Dashed (blue) ($\phi = 2\pi/3$) and dot-dashed (red) ($\phi = 2\pi/3$) lines lie $1\%$ below the unitary line because of the degradation due to the coupling to the environment.}
\label{fig:fidpur_encoding}
\end{figure}

Including non-linearities, losses, and decoherence, we studied the average fidelity and purity decay for arbitrary states
in Fig. 2, where the dependence of these quantities with the azimutal angle $\vartheta$ is shown for two typical values of the
polar $\phi$-angle in the Bloch sphere. For the above parameters we find, see Fig. 2,  average fidelity between  $96.2\%$ (for the states in the equator) and $98.9\%$ (for states in both poles of the Bloch sphere). Upper solid line (in black) corresponds to the ideal average fidelity given by Eq.(\ref{averageF}) where dissipation and dephasing are not considered. In this case, the obtained values are $1\%$ higher than those arising from numerical simulations that include losses, decoherence, and non-linearities. 
 Purity ranges between  $97.3\%$ (in the equator) and $99.3\%$ (in the poles). 

For these parameters the analytic estimation of the rotation angle $\varphi$ appearing in the evolution operator in Eq.(\ref{evol}) is $\varphi = 20.05$ which is the sum of $\Delta t = 20.1$ and the contribution from the dynamical AC-Stark shift given by $1/2 \, \Delta \, t \, (g_d/\Delta)^2 = -0.05$. To obtain a better quality gate the angle $\varphi$ to be compensated can be numerically estimated using an optimization algorithm to maximize the fidelity after the sequence of the operations described above. With this procedure we find that the best value $\varphi = 19.98$ which is very close to the above analytic estimation. 

The imperfection in the compensation angle introduces a systematic error in the gate that is very small. Thus, the maximum achievable fidelity in the absent of loses and decoherence in the equator for the squeezing parameter $r = 1.5$ is $97\%$ (while it would be $97.5\%$ for the perfect unitary controlled-squeeze gate).  The fidelity can be increased not only by raising the squeezing parameter $r$ but also by improving the compensation method discussed above. This can be done, not only by increasing the parameter $\chi$, but also by doing a careful calibration of the duration of each pulse in the experiment.

\section{Conclusions} \label{sec: conclusiones} 

We propose the use of a controlled-squeeze gate in a cQED setup. This is a universal gate when combined with Gaussian operations on the resonator state and single-qubit unitaries. We discuss how to use it to encode quantum states initially stored in the qubit, mapping them onto resonator states that are superpositions of $4n$ or $4n+2$ photon states, thereby making photon losses detectable through simple parity measurements. 
Instead, since both universal gates can be implemented in the architecture, one could choose the most convenient one for the task at hand. For example, the preparation of $|\chi_+\rangle$ states, which, as remarked in \cite{paz2020}, have interesting properties for metrological purposes, is simpler using $\textbf{C-Sqz}$ than using $\textbf{C-Dsp}$. The ideas presented in this paper can be generalized in various ways. For instance, when two modes are present in the resonator, with frequencies $\omega_{\rm A}$ and $\omega_{\rm B}$, a controlled gate that applies a two-mode squeezing operator to the field can be implemented by driving the SQUID with an suitably chosen frequency (the sum of the above mode frequencies, appropriately dressed by the coupling with the qubit). Another possible generalization (which can be used in single- or two-mode states) is to build a gate that applies squeezing operators when the qubit is in state $|0\rangle$, and a different one when the qubit is in state $|1\rangle$ (which can be done by driving the SQUID with two different frequencies and phases). These gates could be useful in reducing the errors caused by an imperfect compensation method (as the encoding studied in the paper required only one application of this gate). These studies will be presented elsewhere.


\section*{Acknowledgments}

This research of NDG, PIV, FCL, and JPP was funded by Agencia Nacional de
Promocion Científica y Tecnológica (ANPCyT), Consejo
Nacional de Investigaciones Cientıficas y Técnicas
(CONICET), and Universidad de Buenos Aires
(UBA). RGC acknowledges support from the Yale Quantum Institute.

\appendix

\section{Universality of $\textbf{C-Sqz}(r,\theta)$ }\label{universalidad}  

In this section, we demonstrate the universality of the controlled-squeeze gate as follows. We will denote it as $\textbf{C-Sqz}(r,\theta)$ as it applies the squeezing operation $\hat{S}(r,\theta)$, conditioned on the state of a control qubit.  The squeezing operator $\hat{S}(r,\theta)$, is defined as 

\begin{equation}\hat{S}(r,\theta) = \exp(\frac{r}{2} (e^{-i \theta} \hat{a}^2 -  e^{i \theta} \hat{a}^{\dagger 2})), \label{s-operator}\end{equation}
and satisfies the following simple property when combined with the displacement operation 

$$\hat{D}(\gamma) = e^{\gamma \hat{a}^\dagger - \gamma^* \hat{a}},$$ 
i.e., 
\begin{equation}\hat{S}(r,\theta) \hat{D}(\gamma) \hat{S}^{-1}(r,\theta) = \hat{D}(\gamma^\prime),\label{compo}\end{equation} where  
$$\gamma^\prime = \gamma\cosh r-\gamma^{*}e^{i\theta}\sinh r,$$ 
where we have also used that 

\[
\hat{S}^\dagger(\zeta) \, \hat{a} \, \hat{S}(\zeta) = \hat{a} \cosh(r) - \hat{a}^\dagger e^{i\theta} \sinh(r),
\]
\[
\hat{S}^\dagger(\zeta) \, \hat{a}^\dagger \, \hat{S}(\zeta) = \hat{a}^\dagger \cosh(r) - \hat{a} e^{-i\theta} \sinh(r).
\]

Thus, applying the displacement $\hat{D}(\gamma)$ in between $\hat{S}^{-1}(r,\theta)$ and $\hat{S}(r,\theta)$ is equivalent to the application of a different displacement $\hat{D}(\gamma^\prime)$. 

The above relation in (\ref{compo}) between squeezing and displacement operators can be simply generalized to the case of control gates by showing that 

\begin{eqnarray}&&\hat{D}^{-1}(\gamma) \textbf{C-Sqz}(r,\theta) \hat{D}(\gamma) (\textbf{C-Sqz})^{-1}(r,\theta) \nonumber \\ &&= \textbf{C-Dsp}(\gamma^\prime - \gamma).\nonumber \end{eqnarray}

\section{Hamiltonian for a SQUID terminated resonator}\label{ap:Hamiltonian}

In this Appendix we will show how to derive, using the main ingredients of cQED theory, the Hamiltonian used to describe the SQUID-terminated superconducting resonator. In order to do this we will start from the classical Lagrangian of the composite system, obtain the Hamiltonian for the modes and finally quantize them. We set units with $\hbar = 1$, and therefore the quantum magnetic flux $\Phi_0 = 2\pi/(2 e)$.

\subsection{Lagrangian formulation}

We start our derivation with the Lagrangian for the dimensionless phase field $\Phi(x,t)$ inside a superconducting resonator of length $d$ with inductance $L_0$ and capacitance $C_0$ per unit length, terminated in a SQUID at $x=d$ \cite{Wustmann2013}. This Lagrangian is simply the sum of the one corresponding to the transmission line which occupies the interval $0\leq x < d$ and the one of the SQUID located at $x = d$:

\begin{eqnarray}
L &=& \left(\frac{1}{2e}\right)^{2}\frac{C_{0}}{2}\int_{0}^{d}(\dot{\Phi}^{2}-v^{2}\Phi'^{2}) \, dx \\
&+& \left(\frac{1}{2e}\right)^{2}2C_{J} \int_{0}^{d} \frac{\dot{\Phi}^{2}}{2} \delta(x- d) \, dx \nonumber \\ 
&+& 2E_{J} \int_{0}^{d} \cos(\Phi)\cos\left(2e \phi(t)\right)\delta (x - d) \, dx , \nonumber \label{LQT}
\end{eqnarray}
where $\dot \Phi$ and $\Phi'$ denote time and spatial derivatives of the phase field, respectively. 
Here $v=1/\sqrt{L_{0}C_{0}}$ is the speed at which the wave propagate in the resonator, whereas $\phi(t)$ is the external magnetic field flux applied to the SQUID.  It is worth noticing that term concentrated in $x = d$ enables the derivation of the equations for the phase $\Phi_d$ for symmetric SQUIDS, with two identical Josephson junctions, each with Josephson energy $E_J$ and capacitance $C_J$. In what follows we will use the notation $\Phi_d(t)=\Phi(d,t)$. 

Evaluating the Euler-Lagrange equation for $\Phi(x,t)$, we obtain 

\begin{equation}
\ddot{\Phi}-v^{2}\Phi^{\prime\prime} + \left(\frac{1}{2e}\right)^{2}2C_{J}\ddot{\Phi}_{d}+2E_{J}\cos\left(2e \phi(t)\right)\sin(\Phi_{d})=0. \label{EL}
\end{equation}

Evaluating previous equation for $x < d$, one obtains the wave equation for the phase field, 

\begin{equation}
\ddot{\Phi}-v^{2}\Phi^{\prime\prime}=0. \label{we}
\end{equation}

The equation for $\Phi_d(t)$ can be obtained after integrating Eq. (\ref{EL}) in the spatial coordinate, between $d - \epsilon$ and $d + \epsilon$, for $\epsilon \rightarrow 0$. While doing this, one has to take into account the existence of a discontinuity in the first spatial derivative of $\Phi(x,t)$ at $x = d$. Thus, the equation for the phase field at the position of the SQUID reads as:

\begin{eqnarray}
\label{eq:borde2}
&\,&\left(\frac{1}{2e}\right)^{2} 2C_{J}\ddot{\Phi}_{d} +  2E_{J}\cos\left(2e \phi(t)\right)\sin(\Phi_{d})
\nonumber \\ &&+ \left(\frac{1}{2e}\right)^{2}C_{0}v^{2}\Phi^\prime_{d}=0. 
\end{eqnarray}
The last two equations along with the Neumann boundary condition at $x = 0$, $(\partial_x\Phi)(0,t)=0$,  define completely the field for the SQUID-terminated resonator.

In order to solve the above equations we first assume the validity of a linear approximation for the phase field $\Phi$ ($\Phi\ll1$), and take into account non-linearities in a perturbative way later.
In this case, when the external flux is constant $\phi(t)=\bar{\phi}$ a basis of solutions for the above equations can be found: 

\begin{equation}
\label{eq:campo2}
\Phi(x,t)=e^{i\omega_{n}t}\psi_{n}(x)
\end{equation}
\begin{equation}
\label{eq:base2}
\psi_{n}(x)=\frac{1}{N_{n}}\cos(k_{n}x),
\end{equation}
where $N_{n}$ is a normalization constant. For these modes to satisfy the wave equation (\ref{we}), the frequency $\omega_{n}$ must be such that $\omega_{n}^2=k_{n}^2v^2$ is the angular frequency for the modes. Moreover, the equation (\ref{eq:borde2}) for $\Phi_d$ is satisfied only if the wave number $k_n$ satisfy the following transcendental equation:  

\begin{eqnarray}
\label{eq:ks2}
&&\, \left(\frac{1}{2e}\right)^{2}C_{0}v^{2}k_{n}\tan(k_{n}d) \\ && =2E_{J}\cos\left(2e\bar{\phi}\right)-\left(\frac{1}{2e}\right)^{2}2C_{J}v^{2}k_{n}^{2}. \nonumber 
\end{eqnarray}

This transcendental equation uniquely determines the spectrum of the SQUID-terminated resonator $\omega_n$ for a given external magnetic flux $\bar\phi$.

It is important to stress that the above spatial modes are orthonormal in the following internal product
\begin{equation}
\label{eq:innprod}
\langle\psi_{i},\psi_{j}\rangle=\int_{0}^{d}\epsilon(x)\psi_{i}(x)\psi_{j}(x)dx,
\end{equation}
where $\epsilon(x)=\left(\frac{1}{2e}\right)^{2}C_{0}+\left(\frac{1}{2e}\right)^{2}2C_{J}\delta(x-d)$. 

We now analyze the system for a time-dependent magnetic flux $\phi(t)=\bar{\phi}+\delta\phi(t)$ that oscillates around a constant value $\bar{\phi}$. In this case we can define spatial modes $\psi_n(t,x)$ with an explicit time dependence arising from the dependence of $k_n$ on time through the generalization of the transcendental Eq.(\ref{eq:ks2}) to the time dependent case. This new basis (which defines the instantaneous spatial modes) is orthonormal in the same inner product defined above. Using this ansatz for the spatial modes, the phase field can be expanded as 
\begin{equation}
\label{eq:expan}
	\Phi(x,t)=\sum_{n}Q_{n}(t)\psi_{n}(t,x),
\end{equation}
where the amplitudes $Q_{n}(t)$ are time dependent functions that contain all the dynamical information of the problem. Clearly, in this basis $\dot\Phi(x,t) = \sum_n(Q_n(t) \dot\psi_n(t,x) + \dot Q_n(t) \psi_n(t,x))$ and therefore we may evaluate, 

\begin{eqnarray}
    \frac{1}{2}\int_0^d  dx \, \epsilon (x) \, \dot\Phi^2 &=& \frac{1}{2} \sum_k \dot Q_k^2 + \frac{1}{2} \sum_{kj} S_{kj}Q_k Q_j \nonumber \\ &-&  \sum_{kj} A_{kj} \dot Q_k Q_j ,\end{eqnarray}
where 

\begin{eqnarray}
    S_{kj} &=& \langle \dot\psi_k , \dot\psi_j\rangle =\int_0^d dx \, \epsilon (x) \, \dot\psi_k \, \dot\psi_j ,\nonumber \\ 
    A_{kj} &=& \langle \psi_k , \dot\psi_j\rangle = \int_0^d dx \, \epsilon (x) \, \psi_k \, \dot\psi_j .\nonumber 
\end{eqnarray} 
Note that, due to the orthogonality of the eigenfunctions, the matrix $A_{kj}$ is antisymmetric.  The matrix $S_{kj}$ is obviously symmetric. They ($A_{ij}$ and $S_{ij})$ are, in short, coupling functions that depend on time. 

A similar calculation can be done for the spatial derivatives (for details on a related calculation 
see the appendix of Ref. \cite{fosco2013})

\begin{eqnarray}
    \int_0^d  \, dx \, \Phi^{\prime 2} &=& \sum_{kj} Q_k \, Q_j \int_0^d \, dx \, \psi_k^\prime \, \psi_j^\prime \\ &=& \sum_{kj} Q_k \, Q_j \int_0^d dx \, \left[\left( \psi_k \, \psi_j^\prime\right)^\prime - \psi_k \, \psi_j^{''} \right] \nonumber \\ &=&  \sum_{kj} Q_k \, Q_j \int_0^d dx \, \left[\left( \psi_k \, \psi_j^\prime\right)^\prime + k_j^2 \, \psi_k \, \psi_j \right],\nonumber \label{rel}
\end{eqnarray}
where we have used the wave equation in the last term. It is worth noticing that because the discontinuity at $x = d$, the transcendental equation implies the following relation

\begin{eqnarray}
    && \int_0^d \, dx \, \left( \psi_k \psi_j^\prime\right)^\prime \, = \\ && 
    \left(\frac{2C_J}{C_0} k_j^2 - \frac{1}{(2e)^2} \frac{2E_J \cos(2e\phi(t))}{C_0 v^2}\right) \psi_k(d) \, \psi_j(d), \nonumber 
\end{eqnarray}
that, when combined with Eq.(\ref{rel}), it allows us to cancel those terms that appear in the Lagrangian (\ref{LQT}), located at $x = d$.

Therefore, using the orthogonality of the  instantaneous modes $\psi_n(t,x)$, defined above, we obtain a Lagrangian for the coefficients $Q_n(t)$ (which play the role of the generalized coordinates for this problem), 

\begin{equation}
L=\frac{1}{2}\sum_{i}\left(\dot{Q_{i}}^{2}-\omega_{i}^{2}Q_{i}^{2}\right)- \sum_{i,j}A_{ij}\dot{Q_{i}}Q_{j}
+\sum_{i,j}S_{ij}Q_{i}Q_{j}, \label{LQn}
\end{equation}
where the frequency in the last equation is time dependent through: $\omega_i = \omega_i(\phi(t)) = v k_i(\phi(t))$ (and $k_i(t)$ are the solutions of the transcendental equation).

\subsection{Hamiltonian formulation}

From the Lagrangian above (Eq.(\ref{LQn})), the classical Hamiltonian can be found by first calculating the canonical conjugated momenta 
\begin{equation}
\label{eq:P}
P_{i}=\frac{\partial L}{\partial\dot{Q}_{i}}=\dot{Q_{i}}-\dot{\delta\phi}\sum_{k}M_{ik}Q_{k},
\end{equation}
where we have used that $\dot\psi_k = d\phi_k/d\phi(t) \, \dot{\delta\phi}$ and therefore we can write $A_{ik} = \dot{\delta\phi} \, M_{ik}$ with $M_{ik} =\langle \psi_i , d\psi_k/d\phi\rangle$. Thus, we can write the new coordinates as, 
\begin{equation}
\dot{Q_{i}}=P_{i}+\dot{\delta\phi}\sum_{k}M_{ik}Q_{k},
\end{equation}
and then, after the Legendre transformation, it is possible to write a Hamiltonian: 
\begin{equation}
    H=\sum_{i}P_{i}\dot{Q}_{i}-L =\frac{1}{2}\sum_{i}\left[P_{i}^{2}+\omega_{i}^{2}Q_{i}^{2}\right]+\dot{\delta\phi}\sum_{i,k}M_{ik}P_{i}Q_{k}. \label{eq:hamiltonian1}
\end{equation}

We can now proceed to quantize the theory by promoting the generalized coordinates and momenta to operators acting on a Hilbert space with canonical commutation rules
\begin{equation}
	[Q_{k},P_{j}]=i\delta_{kj}.
\end{equation}
We can also define annihilation operators as
\begin{equation}
{\tilde a}_{k}=\sqrt{\frac{1}{2{\tilde \omega}_{k}(\phi)}}({\tilde \omega}_{k}(\phi)Q_{k}+iP_{k}),
\end{equation}
which will satisfy the commutation relation $[{\tilde a}_{k},{\tilde a}_{k}^{\dagger}]=1$.

Assuming the system is weakly driven, i.e.,
$|\delta\phi(t)| \ll 1$, the annihilation operators can be approximated by
\begin{equation}
	{\tilde a}_{k}\approx a_{k}+\delta \phi\frac{\omega_{k}^{\prime}}{2\omega_{k}}a_{k}^{\dagger},
\end{equation}
where the annihilation operators correspond to the excitations of the static resonator and $\omega_{k}^{\prime}=(d\omega_{k}/d\phi)(\bar{\phi})$. Finally, replacing these results in Eq.(\ref{eq:hamiltonian1}) we reach to the quantum Hamiltonian

\begin{eqnarray}
\label{eq:Happ}
	H &=& \sum_{k}\left[\omega_{k} \, a_{k}^{\dagger}\, a_{k}
	+\frac{\omega_{k}^{\prime}}{2}\delta \phi \, (a_{k}+a_{k}^{\dagger})^2\right] \\ 
   &+& \delta\dot{\phi}\sum_{i,k}\frac{M_{i k}}{2i}\sqrt{\frac{\omega_{i}}{\omega_{k}}}\left(a_{i}\, a_{k} - a_{i}^{\dagger} \, a_{k} + a_{i} \, a_{k}^{\dagger} - a_{i}^{\dagger}\, a_{k}^{\dagger}\right) , \nonumber 
\end{eqnarray}
where $M_{ij}$ are now coupling constants that depend only on $\bar{\phi}$. 

Here, we have neglected the term associated with the Kerr effect induced by the SQUID, $K_{\rm SQUID}(a^\dagger a^\dagger a a)/2$, since we are working under the assumption that the inductive energy of the resonator is much smaller than the Josephson energy of the SQUID, i.e. (using now the notation introduced in Fig. 1 for physical parameters in the SQUID-terminated resonator), $\sigma=E_{LR}/(2 E_{JR} \cos\left(2e\phi\right))\sim10^{-2}$ (where $E_{LR}=(1/2e)^2(1/L_0d)$) which, as shown in \cite{Wustmann2013}, guarantees that
\begin{align}
    K_{\rm SQUID}/\omega \approx \sigma^3 \frac{\pi Z_0}{2R_q}\leq10^{-5},
\end{align}
where $Z_0=\sqrt{L_0/C_0}$ is the cavity impedance and $R_q=1/(2e^2)$ is the resistance quantum. In other words, under these assumptions, the coefficient associated with the Kerr effect is negligible. We have set $E_{CR}/\hbar = 10 \ \text{GHz}$ and $E_{JR}/\hbar = 9 \ \text{THz}$ (with a typical current on the SQUID of $I\sim 3 \ \mu$A).

Finally, the Hamiltonian of Eq.(\ref{eq:1}) is simply the one-mode approximation of Eq.(\ref{eq:Happ}) capacitively coupled to a qubit in the dispersive regime \cite{Blais2021}. More precisely, we are assuming that the resonator-qubit coupling constant $g_q$, is much smaller than the detuning, $\Delta_{q}=\omega_q-\omega$, making the induced Kerr non-linearity negligible 

\begin{equation}
    K_{\rm qubit}/\omega=- \frac{1}{2} E_{CQ}\left( \frac{g_q}{\Delta_q} \right)^4 \sim 10^{-4},\end{equation} 
    where $g_q=\sqrt{-\chi \Delta_{q}(\Delta_{q}-E_{CQ})/E_{CQ}}$ is the qubit-resonator coupling constant and $E_{CQ}=150 \ \text{MHz}$ is the capacitive energy of the qubit.

Regarding the cross-Kerr we have considered the conditions in which this effect is minimal and can be safely ignored without significantly affecting the system's behavior. It is important to emphasize that in the simulations we performed in Section IV (Fig. 2), we considered the full model, without neglecting any of the non-linear terms.

\section{Encoding protocol}
\label{ap: ep}

In this Appendix we will show how to encode a qubit in a resonator state using the subspace spanned by $|\chi_\pm\rangle$. These states are defined as superpositions of two states squeezed along orthogonal directions. Thus,  

\begin{widetext}

\begin{eqnarray}
       |\chi_{+}\rangle &=& \frac{1}{\sqrt{2}c_+}(|r,\theta\rangle+|r,\theta+\pi\rangle)=\frac{1}{c_{+}\sqrt{2\cosh r}}\sum_{k=0}^{\infty}\frac{\sqrt{(4k)!}}{2^{2k-1}(2k)!}(\tanh r)^{2k}e^{i2k\theta}|4k\rangle,
\\
   |\chi_{-}\rangle &=& \frac{1}{\sqrt{2}c_-}(|r,\theta\rangle-|r,\theta+\pi\rangle)
    =\frac{-1}{c_{-}\sqrt{2\cosh r}}\sum_{k=0}^{\infty}\frac{\sqrt{(4k+2)!}}{2^{2k}(2k+1)!}(-\tanh r)^{2k+1}e^{i(2k+1)\theta}|4k+2\rangle , 
\end{eqnarray}

\end{widetext}
where $c_\pm = \sqrt{1 \pm \frac{1}{\sqrt{\cosh 2r}}}$. 

As it was notice in Sec.\ref{sec:encoding}, $|\chi_\pm\rangle$ are respectively superpositions of states with $4 n$ and $4n+2$ photons. Therefore they remain orthogonal when a photon is lost. Moreover, the lost of a photon is associated with the change of the parity of the states and can be therefore detected by means a simple parity measurement. 

The encoding protocol is composed by the following steps:

\begin{enumerate}
    \item We begin with the resonator in a vacuum state and the control qubit in an arbitrary state:
    \begin{equation} 
    |\psi_{QR}\rangle=(\alpha|0\rangle+\beta|1\rangle) \otimes |0\rangle\nonumber
    \end{equation}
    \item We apply a Hadamard gate on the qubit and obtain:
    \begin{eqnarray}
    |\psi_{QR}\rangle &=&  \frac{1}{\sqrt{2}}\left[\alpha\left(|0\rangle+|1\rangle\right)+\beta\left(|0\rangle-|1\rangle\right)\right]\otimes |0\rangle \nonumber \\ &=& \frac{1}{\sqrt{2}}\left[(\alpha + \beta) |0\rangle + (\alpha - \beta) |1\rangle\right]\otimes |0\rangle \nonumber.
    \end{eqnarray}
\item
    We apply a $\textbf{C-Sqz}(r,\theta+\pi)$ operator. We assume this is an ideal operation that squeezes the states of the resonator only if the state of the qubit is $|1\rangle$. Then, we obtain 

\begin{equation}
    |\psi_{QR}\rangle = \frac{1}{\sqrt{2}}\left[(\alpha + \beta) |0\rangle \otimes|0\rangle + (\alpha - \beta) |1\rangle\otimes |r, \theta+\pi\rangle \right]\nonumber.
    \end{equation}
\item Then we apply a $\pi$-rotation on the qubit that exchanges $|0\rangle \leftrightarrow |1\rangle$ followed a  $\textbf{C-Sqz}(r,\theta)$.  Then obtaining the combined state

\begin{eqnarray}
    |\psi_{QR}\rangle &=& \frac{1}{\sqrt{2}}\left[(\alpha + \beta) \, |1\rangle \, \otimes \, |r,\theta\rangle\right]  \nonumber \\ &+& \frac{1}{\sqrt{2}}\left[(\alpha - \beta) \, |0\rangle \, \otimes \, |r, \theta +\pi\rangle \right]\nonumber.
    \end{eqnarray}
\item Finally we perform an additional $\pi$-rotation on the qubit followed by a Hadamard gate transforming the combined state into

\begin{eqnarray}
    |\psi_{QR}\rangle &=&    \frac{1}{\sqrt{2}}\left[\alpha \, |\chi_ + \rangle+\beta \, |\chi_-\rangle\right] \, \otimes \, |0\rangle \nonumber \\ &+& \frac{1}{\sqrt{2}}\left[\alpha \, |\chi_-\rangle + \beta \, |\chi_+ \rangle\right] \, \otimes \, |1\rangle\nonumber . 
    \end{eqnarray}
\end{enumerate}

\subsection*{Protocol with a non-ideal \textbf{C-Sqz} gate}
\label{niep}

We will now repeat the above protocol when the \textbf{C-Sqz} gate is not ideal but it is the one described in the main text i.e.,  

\begin{equation}
\textbf{ C-Sqz} (r,\theta) = 
\hat S(r,\theta)\otimes |1\rangle\langle1|+ \hat U_0(\varphi) \otimes |0\rangle\langle0|,\label{newcontrol}\end{equation}
where $\hat{S}(r,\theta)$ is the squeezing operator used above, and $\hat U_0(\varphi)$ is the evolution operator of an oscillator with frequency $\tilde \Delta$, which during a time $t$, induces a rotation in phase space in an angle $\varphi = \tilde\Delta t$ (i.e. $\hat U_0 (\varphi)= \exp(-i \tilde\Delta \hat a^\dagger \hat a)$). The new encoding protocol is composed by the following steps:

\begin{enumerate}
    \item We begin with the resonator in a vacuum state and the control qubit in an arbitrary state:
    \begin{equation} 
    |\psi_{QR}\rangle=(\alpha|0\rangle+\beta|1\rangle) \otimes |0\rangle\nonumber
    \end{equation}
    \item We apply a Hadamard gate on the qubit and obtain:
    \begin{equation}
    |\psi_{QR}\rangle = \frac{1}{\sqrt{2}}\left[(\alpha + \beta) |0\rangle + (\alpha - \beta) |1\rangle\right]\otimes |0\rangle \nonumber.
    \end{equation}
\item
    We apply the $\textbf{C-Sqz}(r,\theta+\pi)$ operator defined in (\ref{newcontrol}). This is an operation that squeezes the states of the resonator only if the state of the qubit is $|1\rangle$ and evolves freely if the qubit state is $|0\rangle$. Then, we obtain 

\begin{eqnarray}
    |\tilde\psi_{QR}\rangle &=& \frac{1}{\sqrt{2}}\left[(\alpha + \beta) \, |0\rangle \, \otimes \, \hat U_0(\varphi) |0\rangle \right] \nonumber \\ &+& \frac{1}{\sqrt{2}}\left[ (\alpha - \beta) \, |1\rangle \, \otimes \,  |r, \theta+\pi\rangle \right] \nonumber \\ &=& 
    \frac{1}{\sqrt{2}}\left[(\alpha + \beta) \, |0\rangle \, \otimes \, |0\rangle \right] \nonumber \\ &+&  \frac{1}{\sqrt{2}}\left[ (\alpha - \beta) \, |1\rangle \, \otimes \,  |r, \theta+\pi\rangle \right]\nonumber, 
    \end{eqnarray}
\item Then we apply a $\pi$-rotation on the qubit that exchanges $|0\rangle \leftrightarrow |1\rangle$ followed by another  $\textbf{C-Sqz}(r,\theta)$, obtaining the combined state

\begin{eqnarray}
    |\tilde\psi_{QR}\rangle &=& \frac{1}{\sqrt{2}}\left[(\alpha + \beta) \, |1\rangle \, \otimes \, |r,\theta\rangle \right] \nonumber \\ &+& \frac{1}{\sqrt{2}}\left[ (\alpha - \beta) \, |0\rangle \, \otimes \, \hat U_0(\varphi) \, |r, \theta+\pi\rangle \right] \nonumber.
    \end{eqnarray}
\item In order to compensate the effect of the rotation induced by the operator $\hat U_0(\varphi)$ we proceed as follows:  we turn of the parametric driving (i.e. we set $\epsilon = 0$) and let the system evolve for a time $\tau$ which will be appropriately chosen as described below. Then, the evolution operator (\ref{newcontrol}) induces a rotation in an angle $\Delta \tau$ when the qubit state in $|0\rangle$. Then, after that the total state is 

\begin{eqnarray}
   && |\tilde \psi_{QR}\rangle = \frac{1}{\sqrt{2}}\left[(\alpha + \beta) \, |1\rangle \, \otimes \, |r,\theta\rangle \right] \nonumber \\ &+& \frac{1}{\sqrt{2}}\left[ (\alpha - \beta) \, |0\rangle \, \otimes \, \hat U_0(\Delta \tau) \, \hat U_0(\varphi) \, |r, \theta+\pi\rangle \right] \nonumber.
    \end{eqnarray}
Therefore choosing $\tau$ so that $\Delta \tau + \varphi$ is an integer multiple of $2 \pi$ the effect of the evolution operator $\hat U_0(\Delta \tau) \hat U_0(\varphi)$ can be neglected. After appropriately choosing the angle, the state will satisfies $ |\tilde \psi_{QR}\rangle =  |\psi_{QR}\rangle $.

\item Finally, after the waiting time $\tau$ we perform an additional $\pi$-rotation on the qubit followed by a Hadamard gate obtaining the final state 

\begin{eqnarray}
    |\psi_{QR}\rangle &=&    \frac{1}{\sqrt{2}}|0\rangle \, \otimes  \, \left[\alpha \, |\chi_+ \rangle + \beta \, |\chi_-\rangle\right] \nonumber \\ &+& \frac{1}{\sqrt{2}}|1\rangle \, \otimes \, \left[\alpha \, |\chi_-\rangle + \beta \, |\chi_+\rangle\right] \nonumber . 
    \end{eqnarray}
\end{enumerate}

It is worth noticing that the angle $\varphi$ that needs to be compensated as discussed above can be estimated analytically taking into account the effect of the free rotation and the  AC-Stark shift (that gives a total contribution of $\varphi = 20.05$). 

We numerically solved the Schr\"odinger equation for the resonator in the frame rotating with the qubit frequency $\bar\omega_1$ while the system is parametrically driven with driving frequency $\omega_d = 2 \bar\omega_1$. We used this in order to find the resonator states after the steps 3 and 4 from the above protocol. Using this, we numerically estimated the angle $\varphi$ that needs to be compensated which turned out to be $\varphi = 19.98$, which is rather close to the one estimated analytically. In the numerical simulation we also included the effect of Kerr non-linearities and finally, we used this same strategy to include the effect of losses and decoherence using the master equation shown in Sec. \ref{sec:implemantation}.

\section{Simulation parameters}\label{parametros}

In connection with the setup depicted in Fig. 1, it is important to mention that the numerical simulations are based on the physical parameters used in \cite{Wustmann2013}. These imply that \( \sigma = E_{LR}/(2E_{JR} \cos(2e\phi)) \sim 10^{-2} \) (where \( E_{LR} = (1/2e)^2 1/(L_0 d) \)); with \( \sigma^3 \pi Z_0 2 R_q \leq 10^{-5} \), and \( Z_0 = \sqrt{L_0 / C_0} \) is the cavity impedance and \( R_q = \frac{1}{2e^2} \) is the resistance quantum. Additionally, we set \( E_{CR} / \hbar = 10 \ \text{GHz} \) and \( E_{JR} / \hbar = 9 \ \text{THz} \) (which is consistent with the energy scales involved in the dynamics of superconducting devices like SQUIDs with $I_c= 3 \ \mu$A).

The resonator frequency is chosen as \( \omega / (2\pi) = 6 \ \text{GHz} \), the qubit frequency as \( \omega_q / (2\pi) = 4 \ \text{GHz} \), the driving coupling as \( \bar{g}_d = 50 \ \text{MHz} \), and the driving amplitude \( \epsilon = 0.15 \), which results in \( g_d = 7.5 \ \text{MHz} \). The qubit coupling strength is set to \( \chi / (2\pi) = 8 \ \text{MHz} \). The controlled-squeeze gate is applied for 200 ns, leading to a squeezing parameter \( r \sim 1.5 \).

Losses and decoherence were modeled by accounting for thermal contact between the qubit-resonator system and a bath at 60 mK. For the relaxation times: a qubit relaxation time \( \tau_q = 200 \ \mu \text{s} \), a resonator damping time \( \tau_r = 200 \ \mu \text{s} \), and a qubit dephasing time \( \tau_\varphi = 10 \ \mu \text{s} \).

Finally, the time evolution was simulated using QuTiP, with the resonator Fock space truncated to a basis of dimension \( N = 90 \). With these parameters and assuming a qubit capacitive energy \( E_{CQ} = 150 \ \text{MHz} \), the induced Kerr nonlinearity is three orders of magnitude smaller than the squeezing term, rendering it negligible.

\bibliography{bib}

\end{document}